\documentstyle[twoside,fleqn,espcrc2]{article}

\title{ \vspace{-2.8cm}
\begin{flushright}
{\normalsize HU-TFT-96-32}
\end{flushright}
\vspace{1.5cm}
The order of the phase transition in 3d U(1)+Higgs theory}
\author{M. Karjalainen\address{Department of Physics, PO Box 9, FIN-00014
University of Helsinki, Finland}, M. Laine\address{Institut 
f\"ur Theoretische Physik, 
Philosophenweg 16, D-69120 Heidelberg, Germany} and J. Peisa$^{\rm a}$} 

\begin{document}

\begin{abstract}
We study the order of the phase transition in the 3d U(1)+Higgs
theory, which is the Ginzburg-Landau theory of superconductivity. 
We confirm that for small scalar self-coupling the transition 
is of first order. For large scalar self-coupling the transition
ceases to be of first order, and a non-vanishing scalar mass suggests 
that the transition may even be of higher than second order.  

\end{abstract}

\maketitle

\input epsf

\section{Introduction}

Scalar electrodynamics (i.e., the U(1)$+$Higgs \linebreak 
theory) in three dimensions
describes several different physical systems. First, it is an 
effective high-temperature theory for 4d scalar electrodynamics 
(with or without fermions)~\cite{a}. 
In particular, the 3d theory can be used
to study the finite temperature phase transition in the 4d theory  
for small 4d coupling constants. Second, it is directly 
the Ginzburg-Landau theory of superconductivity. 
In this case, as well, the properties of the phase transition 
between the normal and superconducting states 
are of much interest~\cite{hlm,dh,b,m,bph,w}. 

In general, the properties of the phase transition in 
the 3d U(1)+Higgs theory cannot be reliably studied in perturbation
theory. This is due to the fact that the 3d coupling constants  
are dimensionful, resulting in infrared problems if there
are nearly massless excitations (the true dimensionless 
expansion parameter is a coupling constant divided by a mass 
$\sim e_3^2/m_H(T)$).
While the IR-problems are not quite as severe as in 
non-abelian theories containing gauge self-interactions, 
they nevertheless exist. In addition,  
non-perturbative topological defects 
play a role in U(1)$+$Higgs. % theory (see, e.g., \cite{br}).

Due to the difficulties mentioned, the 3d \linebreak
U(1)$+$Higgs theory should be 
studied non-perturbatively~\cite{hlm,dh,b,m,bph,w}. 
More specifically, some important
problems to be addressed are: (a)
the existence and order of a phase transition as a 
function of the scalar self-coupling, (b)
the characteristics of the phase transition: 
the critical temperature $T_c$, latent heat,
surface tension, correlation lengths in the two phases, and
(c) the convergence of perturbation theory in the broken phase.
We present here preliminary lattice results on (a); 
more complete results on (a) and (b)
will be presented elsewhere~\cite{kklp}. (c)
was addressed in~\cite{kp}. 

\section{The action and the parameters}

The continuum U(1)$+$Higgs theory in 3d is defined by the action 
\begin{eqnarray}
S &=& \int d^3x \bigg[
\frac{1}{4}F_{ij}F_{ij} + (D_i\phi)^*(D_i\phi) \nonumber \\
&& + m_3^2 \phi^*\phi + \lambda_3 \left(\phi^*\phi\right)^2\bigg], \label{ac}
\end{eqnarray}
where $D_i=\partial_i + i e_3 A_i$.
The scale of the theory is given by 
the dimensionful [GeV] gauge coupling $e_3^2$, 
and the other two parameters are dimensionless,   
\[
x= \frac{\lambda_3}{e_3^2}, \,\,\, y=\frac{m_3^2(e_3^2
)}{e_3^4}.
\] 
The parameter $x$ is roughly
proportional to the ratio of the squares of the scalar and vector
masses deep in the broken phase, $x\sim m_H^2/m_W^2$,
while $y$ is related to temperature,
$y \sim (T-T_c)/T_c$. Here $T_c$ is the tree level 
critical temperature.

The Ginzburg-Landau theory of superconductivity 
is defined just by eq.~(\ref{ac}). 
Conventionally, $m_3^2, \lambda_3$ 
are denoted by $a, b$ in that context,
and $x$ is represented by the Ginzburg parameter
$\kappa=\sqrt{x}$. 
If $\kappa < 1/\sqrt{2}$ the superconductor is of type I, if
$\kappa > 1/\sqrt{2}$ it is of type II. 
The phase transition is %believed to be 
of first order for strongly type I superconductors,
and for strongly type II superconductors it is usually 
assumed to be of second order.

\section{Discretization and simulations}

The action in eq.~(\ref{ac})
can be discretized in the usual way. We use the compact
formulation for the gauge field, so that the 3d  
lattice action is
\begin{eqnarray}
S &=& \beta_G \sum_P {\mbox{Re}}\, (1-U_P) \nonumber \\
  & & -\beta_H \sum_{x, i} {\mbox{Re}}\, 
\phi^*(x) U_i(x)\phi(x+\hat{i}) \nonumber \\   
  & & +\sum_{x} \phi^* \phi + \sum_{x} \beta_R 
		\left[1-\phi^*(x)\phi(x)\right]^2. \label{laq}
\end{eqnarray}
Here $U_i(x)$ is the compact link variable,  $U_P$ is the product of
link variables around a plaquette $P$ and $\phi$ is a complex scalar
field located on sites. It would also be possible to use a 
non-compact formulation~\cite{dh,b}. 

The three coupling constants in the lattice action are related to
continuum parameters through a {\em constant physics curve}, which can
be calculated using lattice perturbation theory. Due to the fact that  
the couplings are dimensionful, the relation can be found exactly 
with a 2-loop calculation~\cite{ct}.
The relevant relations are
\begin{eqnarray*}
\beta_G = {1\over a e_3^2}, && \beta_R = {x \beta_H^2\over 4 \beta_G},
\end{eqnarray*}
\begin{eqnarray*}
2 \beta_G^2\left({1\over\beta_H} - 3 - 
	 {2\beta_R\over \beta_H}\right) &=&
 y -(2+4x){\Sigma\beta_G\over 4 \pi}  \\ 
&&\hspace{-2cm} -{1\over 16\pi^2}\bigg[\left(-4+8x-8x^2\right) \\
&&\hspace{-2cm} \times \left(\ln 6\beta_G + 0.09\right) + 25.5 + 4.6x\bigg].
\end{eqnarray*}
The continuum limit is hence at $(\beta_G,\beta_H,\beta_R)=(\infty,1/3,0)$.
Note that increasing $\beta_H$ means decreasing $y$.

The simulations were done using Cray C94 at the Center for Scientific
Computing in Helsinki. 

\begin{figure}
  \centering
  \leavevmode
  \epsfxsize = 7.5cm
  \epsfbox{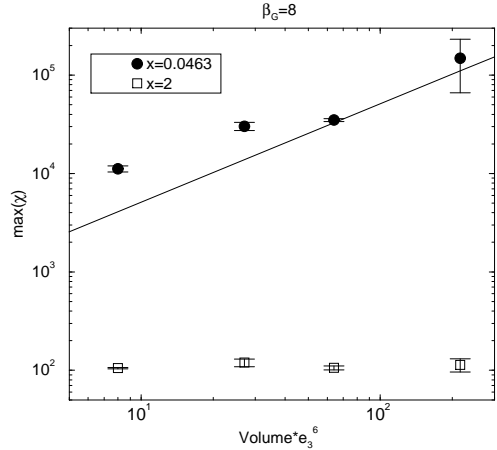}

  \vspace*{-1cm}

  \caption{\protect The maximum of the 
           susceptibility $\chi$ as a function of
	   volume. The straight line indicates the 
	   linear behaviour characteristic of first order
	   transitions. The data point corresponding to the largest volume
	   for $x=0.0463$ was obtained by reweighting the data.}
  \label{scplot1}

  \vspace*{-0.3cm}	
\end{figure}

\section{The order of the transition}

For small $x$, 
the phase transition in the 3d U(1)$+$Higgs theory
is of the first order~\cite{hlm}. 
For larger $x$
(i.e., for type II superconductors), the IR-problems
are more severe since the transition gets weaker, 
and the order has remained unclear.
Arguments in favour of a second order
transition have been given, e.g., in~\cite{dh,b,m,w}, 
but a still higher-order transition
cannot at present be excluded. In particular, 
in the lattice studies in~\cite{dh,b}, the
correlation lengths were not  measured. 

It is interesting to compare the situation  
with that in the 3d SU(2)$+$Higgs theory. There the line
of first-order transitions ends at $x\sim 1/8$, and
for $x > 1/8$ the transition is of higher than 
second order~\cite{mH}. An important difference
between the U(1) and SU(2) cases is that in the latter
all the excitations in the symmetric phase are massive, 
whereas in U(1) there should always be a massless photon
in the continuum limit~\cite{bph,kre,h,bip}.
Consequently, one can define an order parameter for
the continuum theory~\cite{kre}. Away from the continuum limit an 
exponentially small non-perturbative mass is generated
in the compact lattice formulation~\cite{p}, and the phases
are analytically connected~\cite{brfs}.

To study the order of the phase transition, 
we have used two different values
of $x$. The first case $x=0.0463$ corresponds to a strongly type I
superconductor, and the other case $x=2$ to a strongly type II
superconductor.

\begin{figure}
  \centering
  \leavevmode
  \epsfxsize = 7.5cm
  \epsfbox{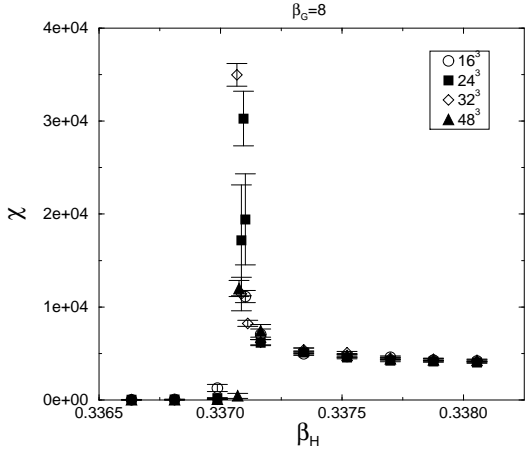}

  \vspace*{-1cm}

  \caption{\protect Susceptibility at $x=0.0463$.}
  \label{scplot2}

  \vspace*{-0.3cm}	
\end{figure}

In the analysis we follow closely~\cite{mH}.
The method employed is to plot the
maximum of the susceptibility as a function of the lattice volume. 
The susceptibility $\chi$ is defined as
\[
\chi = e_3^2 V \left\langle (\phi^*\phi
     - \langle \phi^*\phi\rangle)^2\right\rangle.
\]
In a first order transition $\chi$ grows as the volume
$V$ and in a second order
transition the expected behaviour is $\sim V^\kappa$, $\kappa$
being a critical exponent. If $\kappa\le 0$, or if the transition
is of higher than second order, $\chi \sim V^0$.

\begin{figure}
  \centering
  \leavevmode
  \epsfxsize = 7.5cm
  \epsfbox{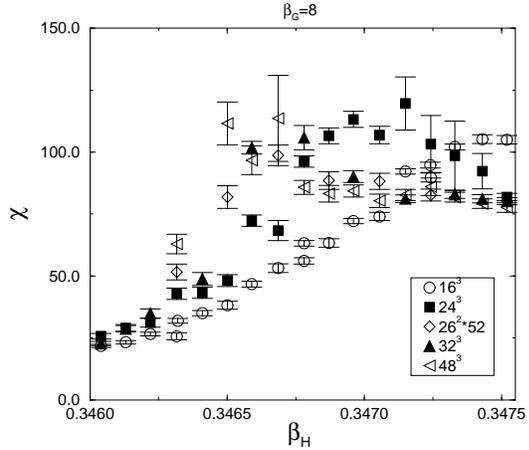}

  \vspace*{-1cm}

  \caption{\protect Susceptibility at $x=2$.}
  \label{scplot3}

  \vspace*{-0.3cm}	
\end{figure}

Our results are plotted in Fig.~\ref{scplot1}, and the individual
susceptibilities in 
Figs.~\ref{scplot2}-\ref{scplot3}. The 
system has a first order phase transition 
at $x=0.0463$, as the maximum of
susceptibility grows linearly with volume. This is
not the case with the $x=2$ data, so the transition in strongly type
II supercondutors is not of first order. The data would
suggest that if the transition is of second order the critical
exponent $\kappa$ is close to zero, as noticed already
in \cite{dh,b}. However, the transition might
also be of higher order.

To study the transition at $x=2$ in more detail, 
we have measured the masses of the scalar and vector excitations.
The masses are measured from the exponential decay of suitably chosen
correlators. For the scalar mass we used the operator
\[
\phi^{\ast}(x)\phi(x),
\]
and for the vector mass
\[
\mathop{\rm Im}\phi^{\ast}(x)U_i(x)\phi(x+\hat{i}).
\]
The photon mass in the symmetric phase can be measured from the imaginary
part of the pla\-quette variable at non-zero momenta~\cite{pho}.

\begin{figure}
  \centering
  \leavevmode
  \epsfxsize = 7.5cm
  \epsfbox{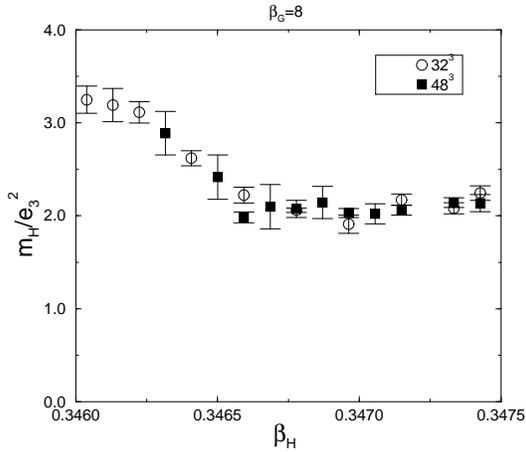}

  \vspace*{-1cm}

  \caption{\protect Scalar mass near the 
    pseudocritical coupling $\beta_{H,c}$, for $x=2$.
    The values of $\beta_{H,c}$ can be seen from the maxima
    of $\chi$ in Fig.~3.}
  \label{smass}

  \vspace*{-0.3cm}	
\end{figure}

\begin{figure}
  \centering
  \leavevmode
  \epsfxsize = 7.5cm
  \epsfbox{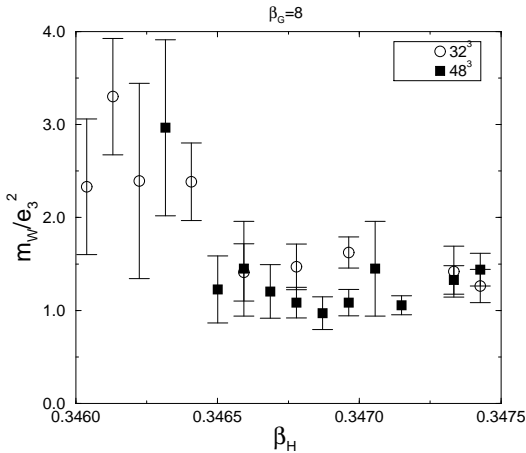}

  \vspace*{-1cm}

  \caption{\protect Vector mass near $\beta_{H,c}$
  for $x=2$. Note that the high-temperature symmetric phase 
  corresponds to $\beta_H < \beta_{H,c}$.}
  \label{vmass}

  \vspace*{-0.3cm}	
\end{figure}

The scalar mass is shown in Fig.~\ref{smass} and
the vector mass in Fig.~\ref{vmass}. The scalar mass
is clearly non-vanishing. For the vector mass the signal 
gets very noisy in the symmetric phase ($y>y_c$, $\beta_H<\beta_{H,c}$), 
and the techniques
of~\cite{ptw} should be used. In general, improved techniques 
make the masses smaller so that, for example,  the vector
mass could go to zero at the phase transition point.
Our very preliminary photon mass measurements in the 
symmetric phase are consistent with zero.
 
In conclusion, we have seen that in type II superconductors 
the phase transition is not of first order. It may even not
be of second order, since we see a non-vanishing scalar
mass. For definite conclusions we will have to
improve the quality of mass measurements and study
the finite $a$ effects.


\begin{thebibliography}{99}

\bibitem{a}
P. Arnold, Phys.\ Rev.\ D 46 (1992) 2628;
K. Farakos, K. Kajantie, K. Rummukainen and M. Shaposhnikov, 
Nucl.\ Phys.\ B 425 (1994) 67;
A. Jakov\'ac, A. Patk\'os and P. Petreczky, 
Phys.\ Lett.\ B 367 (1996) 283.

\bibitem{hlm}
B.I. Halperin, T.C. Lubensky and S.-K. Ma, 
Phys.\ Rev.\ Lett.\ 32 (1974) 292.

\bibitem{dh} 
C. Dasgupta and B.I. Halperin, 
Phys.\ Rev.\ Lett.\ 47 (1981) 1556.

\bibitem{b} 
J. Bartholomew, Phys.\ Rev.\ B 28 (1983) 5378.

\bibitem{m}
J. March-Russell,
Phys.\ Lett.\ B 296 (1992) 364.

\bibitem{bph}
W. Buchm\"uller and O. Philipsen, 
Phys.\ Lett.\ B 354 (1995) 403.

\bibitem{w}
B. Bergerhoff, F. Freire, D.F. Litim, S. Lola and C. Wetterich, 
Phys.\ Rev.\ B 53 (1996) 5734.

\bibitem{kklp}
K. Kajantie, M. Karjalainen, M. Laine and J. Peisa, 
in preparation.

\bibitem{kp}
M. Karjalainen and J. Peisa, 
HU-TFT-96-27 [hep-lat/9607023].

\bibitem{ct}
M. Laine, Nucl.\ Phys.\ B 451 (1995) 484.

\bibitem{mH}
K. Kajantie, M. Laine, K. Rummukainen and M. Shaposhnikov, 
CERN-TH/96-126 [hep-ph/9605288].

\bibitem{kre}
A. Kovner, B. Rosenstein and D. Eliezer, 
Nucl.\ Phys.\ B 350 (1991) 325.

\bibitem{h}
A. Hebecker, Z.\ Phys.\ C 60 (1993) 271.

\bibitem{bip}
J.-P. Blaizot, E. Iancu and R.R. Parwani, 
Phys.\ Rev.\ D 52 (1995) 2543.

\bibitem{p}
A.M. Polyakov, 
Phys.\ Lett.\ B 59 (1975) 82.

\bibitem{brfs}
T. Banks and E. Rabinovici, Nucl.\ Phys.\ B 160 (1979) 349;
E. Fradkin and S. Shenker, Phys.\ Rev.\ D 19 (1979) 3682.

\bibitem{pho}
B. Berg and C. Panagiotakopoulos, Phys.\ Rev.\ Lett.\ 52 (1984) 94;
H.G. Evertz, K. Jansen, J. Jers\'ak, C.B. Lang and T. Neuhaus,
Nucl.\ Phys.\ B 285 (1987) 590.

\bibitem{ptw}
O. Philipsen, M. Teper and H. Wittig, Nucl.\ Phys.\ B 469 (1996) 445.

\end{thebibliography}
\end{document}